\pdfoutput=1
\documentclass[sigconf,nonacm]{acmart}

\graphicspath{{./figures/}}
\usepackage[a-1b]{pdfx} 
\usepackage{array}
\usepackage{amsmath}
\usepackage{algorithmic}
\usepackage[ruled,vlined]{algorithm2e}
\usepackage{pgfplots,pgfplotstable}  
\usepackage{booktabs}
\usepackage{balance}
\usepackage{multirow}
\usepackage{subcaption}
\pgfplotsset{compat=newest}
\usepackage{balance}
\usepackage{changepage} 
\usepackage{tcolorbox}  
\usepackage{caption}
\usepackage{geometry}
\usepackage{longtable}
\usepackage{graphicx}  
\usepackage{multicol}

\renewcommand{\shortauthors}{Cahoon et al.}

\newcommand{\am}[1]{\textcolor{blue}{[Andreas: #1]}}

\newcommand\vldbavailabilityurl{}

\newcommand{\eat}[1]{}
\newcommand{\sysname}{TREX\xspace}
\newcommand{\hotpot}{HotPotQA\xspace}
\newcommand{\msmarco}{MSMarco\xspace}

\newcounter{enum}

\begin{document}
\title{Optimizing open-domain question answering with graph-based retrieval augmented generation}

\begin{abstract}
In this work, we benchmark various graph-based retrieval-augmented generation (RAG) systems across a broad spectrum of query types, including OLTP-style (fact-based) and OLAP-style (thematic) queries, to address the complex demands of open-domain question answering (QA). Traditional RAG methods often fall short in handling nuanced, multi-document synthesis tasks. By structuring knowledge as graphs, we can facilitate the retrieval of context that captures greater semantic depth and enhances language model operations. We explore graph-based RAG methodologies and introduce \sysname, a novel, cost-effective alternative that combines graph-based and vector-based retrieval techniques. Our benchmarking across four diverse datasets highlights the strengths of different RAG methodologies, demonstrates TREX's ability to handle multiple open-domain QA types, and reveals the limitations of current evaluation methods.

In a real-world technical support case study, we demonstrate how TREX solutions can surpass conventional vector-based RAG in efficiently synthesizing data from heterogeneous sources. Our findings underscore the potential of augmenting large language models with advanced retrieval and orchestration capabilities, advancing scalable, graph-based AI solutions.
\end{abstract}

\settopmatter{authorsperrow=3}
\author{Joyce Cahoon}
\email{jcahoon@microsoft.com} \vspace{0.5em}
\affiliation{%
  \institution{Microsoft}
  \city{Redmond}
  \state{WA}
  \country{USA}
}
\author{Prerna Singh}
\email{prernasingh@microsoft.com}
\affiliation{%
  \institution{Microsoft}
  \city{Redmond}
  \state{WA}
  \country{USA}
}
\author{Nick Litombe}
\email{nicklitombe@microsoft.com}
\affiliation{%
  \institution{Microsoft}
  \city{Austin}
  \state{TX}
  \country{USA}
}
\author{Jonathan Larson}
\email{jolarso@microsoft.com}
\affiliation{%
  \institution{Microsoft}
  \city{Redmond}
  \state{WA}
  \country{USA}
}
\author{Ha Trinh}
\email{trinhha@microsoft.com}
\affiliation{%
  \institution{Microsoft}
  \city{Redmond}
  \state{WA}
  \country{USA}
}
\author{Yiwen Zhu}
\email{yiwzh@microsoft.com}
\affiliation{%
  \institution{Microsoft}
  \city{Mountain View}
  \state{CA}
  \country{USA}
}
\author{Andreas Mueller}
\email{amueller@microsoft.com}
\affiliation{%
  \institution{Microsoft}
  \city{Mountain View}
  \state{CA}
  \country{USA}
}
\author{Fotis Psallidas}
\email{fotis.psallidas@microsoft.com}
\affiliation{%
  \institution{Microsoft}
  \city{New York}
  \state{NY}
  \country{USA}
}
\author{Carlo Curino}
\email{carlo.curino@microsoft.com}
\affiliation{%
  \institution{Microsoft}
  \city{Redmond}
  \state{WA}
  \country{USA}
}
\renewcommand{\shortauthors}{Cahoon et al.}

\maketitle



\section{Introduction}
\label{sec:intro}

Traditionally, knowledge workers---such as executives, managers, and analysts---relied on data warehousing and operational databases to make faster, more informed decisions \cite{chaudhuri1997overview}. Today, the range of decision support tools has expanded to include foundational models, with AI software spending projected to reach \$297.9 billion by 2027 \cite{gartner_ai_software_forecast}. Although the modern knowledge worker operates in a vastly different environment, one not limited to carefully curated data warehouses with complex multi-dimensional models for online analytical processing (OLAP) or highly structured operational databases supporting online transaction processing (OLTP) \cite{chaudhuri1997overview}, the core challenges and types of questions that these decision support systems address remain highly relevant.

The need for decision support tools capable of processing vast amounts of unstructured data at scale is particularly evident in open-domain question answering (QA). This field, which spans Natural Language Processing (NLP), Information Extraction (IE), and Information Retrieval (IR), focuses on answering questions without relying on predefined context \cite{zhu2021retrieving}. Large language models (LLMs) such as GPT \cite{openai2024gpt4technicalreport}, Claude \cite{TheC3}, and Llama \cite{dubey2024llama3herdmodels} have emerged as powerful tools for this purpose, generating human-like responses to complex queries while processing extensive text inputs. In various enterprise applications that leverage LLMs, user queries can often be classified as either OLTP or OLAP. OLTP-style queries are simple, fact-based questions that can be answered through direct key-value lookups, retrieval from single text snippets, or by locally traversing multiple related pieces of text, while OLAP-style queries are open-ended, thematic, and require aggregating, synthesizing and abstracting information across multiple documents \cite{yu2024evaluation}. Just as operational databases are optimized for OLTP tasks and data warehouses for OLAP workloads, specialized LLM applications are now emerging to address these distinct query types, with tailored approaches for both OLTP-like and OLAP-like QA tasks.

Retrieval-augmented generation (RAG) has become a widely adopted LLM-based approach for open-domain QA, leveraging various retrieval strategies that adapt to user input queries. In many commercial applications today, the typical RAG workflow involves indexing large volumes of text, using either inverted indices or dense vector encodings, and then retrieving the most relevant information, which often includes incorporating re-rankers to refine the retrieved documents \cite{gao2024retrievalaugmentedgenerationlargelanguage, chen2023benchmarkinglargelanguagemodels}. 
This retrieved context is subsequently used as input for the generation component, which often involves a language model to generate a response. This straightforward yet powerful methodology is effective for extractive tasks or OLTP-style applications, where answers can be located within a single text snippet embedding---in other words, analogous to a database scenario where a simple key-value lookup suffices \cite{yu2024evaluation}.

However, in OLAP-style settings---where queries are open-ended and require synthesizing information from multiple documents (e.g., ``How is artificial intelligence impacting global job markets?'' or ``What are the latest technology trends?'')---existing RAG systems still struggle to retrieve the most relevant information.
Although recent advancements in hardware and algorithms have increased the context lengths that models can process \cite{dao2022flashattention}, the importance of precise context selection remains. Issues such as context stuffing \cite{llamaindex_long_context_rag}, the ``lost in the middle'' phenomenon \cite{liu2024lost}, and diminishing model performance with singular 
context persists \cite{sun2021long}. While some models \cite{team2024gemini} achieve both high recall and precision, using such long-context modalities can be costly and slow, underscoring the ongoing need for careful selection of what information is included in the context.

Building on these challenges, graph-based RAG solutions such as GraphRAG \cite{graphrag} have been developed to improve retrieval for OLAP-like queries by structuring and synthesizing information across multiple documents. Since graph-based RAG solutions are primarily designed for OLAP-style queries, subsequent advancements, including \citet{sarmah2024hybridragintegratingknowledgegraphs} and \citet{cosmosaigraph}, have introduced query orchestrators to dynamically route questions to the appropriate RAG methodology, ensuring effective handling of all query types. However, misrouted queries \citep{sarmah2024hybridragintegratingknowledgegraphs} and the high cost of maintaining complex graph-based retrieval systems present significant limitations. To mitigate these issues, we introduce TREX (\underline{T}runcated \underline{R}APTOR \underline{E}xpanded Inde\underline{X}), a cost-effective alternative that eliminates the need for expensive orchestrators while reducing the overhead of full knowledge graph indexing \cite{raptor}. This paper evaluates the strengths and limitations of advanced RAG methodologies in open-domain QA and introduces \sysname, a scalable approach that balances semantic depth and computational efficiency. We benchmark \sysname on four diverse datasets representative of enterprise-scale LLM applications to highlight scenarios where our method surpasses existing solutions.


\subsection{Industry Trends}

\begin{figure}
    \centering
    \includegraphics[width=0.35\textwidth]{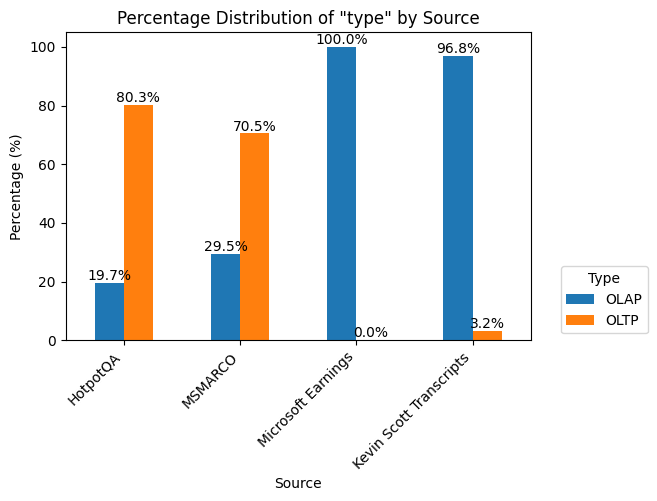}
    \vspace{-.1cm}
    \caption{Breakdown of the percentage of OLTP vs OLAP-style queries within each benchmark.}
    \label{fig:spider_map}
\end{figure}

With our diverse range of customer engagements across sectors such as manufacturing, healthcare, finance and telecommunications, we have identified two distinct classes of queries that are of particular interest to our clients. These query types can be broadly categorized based on their question scope (e.g., local versus global) \cite{graphrag}:

\begin{itemize}
    \item Online Transaction Processing (OLTP): These are straightforward, fact-based queries that can be answered through direct key-value lookups, extraction of specific text snippets, or by locally traversing closely related pieces of information, such as ``What is the price of Microsoft stock today?'' or ``What was Microsoft’s revenue in the fourth quarter of FY2024?'' Such queries are data-driven and generally limited in scope \cite{microsoft_earnings_fy24_q4}.
    \item Online Analytical Processing (OLAP): These queries are open-ended and thematic, requiring the aggregation, synthesis, and abstraction of information across multiple documents. Examples include questions like “What are the key trends in the technology sector?” or “How is the balance between innovation and ethics discussed?” \cite{graphrag} OLAP-style queries are activity-driven and broader in scope, making their evaluation particularly challenging. Unlike fact-based queries, which have clear-cut answers, OLAP queries often yield a diverse range of valid responses, further complicating the assessment of correctness.
\end{itemize}
Insights from our customer engagements reveal that the distribution of queries between OLTP and OLAP-style are quite mixed, with some customers that are exclusively interested in broader, activity-driven, OLAP-type, while others are primarily interested in local, data-driven, OLTP-style queries on their data. While ongoing focus groups need to be conducted to better quantify this distribution, recognizing these query types has directly shaped our selection of four benchmarking datasets, as illustrated by Figure~\ref{fig:spider_map}.

Through our engagements, we have gained valuable insights into the types of workloads and queries that customers expect our 
LLM applications to handle, as well as the pressing need for improved methods to track and assess performance regressions. Yet, evaluating RAG systems remains a significant challenge due to the absence of standardized approaches for open-domain QA. Current evaluation methods are largely benchmark-driven; for example, datasets like HotPotQA---a widely used multi-hop QA benchmark derived from Wikipedia---include human-annotated ground-truth answers, allowing for direct accuracy measurement by comparing model responses against verified correct answers \cite{hotpot}. However, other datasets, such as the Kevin Scott Podcasts \cite{graphrag}, introduce a different challenge---open-ended questions (e.g., ``What is happening in the technology sector?'') where multiple responses are valid. This highlights a critical gap in the industry: the need for robust evaluation frameworks and best practices to consistently judge the quality and correctness of open-domain QA systems \cite{chen2023benchmarkinglargelanguagemodels}. Developing standardized metrics and methodologies will be essential to ensuring reliable and meaningful performance assessments across diverse QA applications.

\subsection{Our Contributions}

We examine two prominent graph-based RAG methodologies, RAPTOR \cite{raptor} and GraphRAG \cite{graphrag}, popularized by Neo4J \cite{neo4j} and LlamaIndex \cite{llamahub_raptor_retriever}, and introduce \sysname, a novel and cost-effective alternative that captures semantic depth while preserving coherence within each text chunk. Our key contributions are as follows:

\begin{itemize}
\item We develop \sysname, which provides a balanced cost-performance solution suitable for both OLAP and OLTP query types.
\item We benchmark graph-based RAG techniques to identify best practices and optimal scenarios for their application, providing insights into when each approach is most effective.
\item We introduce new metrics for evaluating faithfulness, showing where our method surpasses existing approaches, and report comparative win rates for a clearer understanding of each methodology's relative performance.
\item Finally, we demonstrate \sysname's effectiveness in a real-world technical support setting, showcasing its utility with unstructured customer support data.
\end{itemize}

Our extensive benchmarking identifies three key challenges in leveraging LLMs for open-domain question answering: (i) retrieval quality is critical to final performance and presents substantial room for improvement; (ii) state-of-the-art LLMs continue to face difficulties in synthesizing multiple documents without being influenced by irrelevant information; and (iii) evaluating RAG systems remains challenging due to the lack of a standardized approach for open-domain QA. 

These challenges pose promising research directions for developing better systems integrating retrieval and LLMs.

\section{Related Work}\label{sec:related_work}

Text summarization is a foundational task in open-domain QA, with methods broadly categorized as extractive (selecting key sentences) and abstractive (generating new sentences). Hybrid approaches blend both techniques \cite{el2021automatic}. Summarization supports both OLTP-style and OLAP-style queries by condensing large text volumes into accessible gists. While extractive methods suffice for fact-based OLTP queries, OLAP-style queries, requiring synthesis across multiple documents, benefit from recursive and hybrid summarization approaches. Advances in deep learning, particularly large language models (LLMs), have blurred the line between extractive and abstractive summarization, enabling high-quality hierarchical summaries for complex QA tasks \cite{li2024graphreader}.

\subsection{Retrieval-Augmented Generation (RAG)}

Despite improvements in handling long contexts, LLMs still struggle with efficiently processing extended text \cite{liu2024lost}, making RAG essential for supplementing generation with retrieved context. Standard RAG retrieves text chunks from databases, optimized using strategies such as chunking \cite{langchain_github}, entity-based organization \cite{dejong2022mentionmemoryincorporatingtextual}, and token-level granularity \cite{khandelwal2020generalizationmemorizationnearestneighbor}. Retrieval methods range from classical BM25 \cite{robertson2009probabilistic} to supervised and heuristic-based ranking approaches \cite{zhang2022adversarialretrieverrankerdensetext, sachan2023questionsneedtraindense, cormack2009reciprocal, microsoft_hybrid_search_ranking}.

\subsection{Graph-based RAG}
Traditional retrieval methods rely on unstructured text, which limits their ability to capture complex relationships. Graph-enhanced RAG techniques address this by incorporating structured knowledge, improving interpretability and reasoning capabilities. Surveys such as \citet{jin2024large} and \citet{Pan_2024} categorize LLM-graph integrations, distinguishing between text-attributed graphs and knowledge graphs (KGs). Recent approaches leverage graphs to optimize retrieval in various ways. MemWalker \cite{chen2023walking} organizes documents into a hierarchical tree and employs iterative prompting to navigate this structure for more efficient context retrieval. PEARL \cite{sun2023pearl} decomposes complex reasoning tasks into structured sequences of actions, such as identifying participants, summarizing conversations, and executing planned steps. ReadAgent \cite{lee2024humaninspiredreadingagentgist} condenses documents into gists while maintaining a memory directory to facilitate recall of relevant details for task completion. Knowledge Graph Prompting (KGP) \cite{wang2024knowledge} structures documents as a graph and utilizes a traversal agent to retrieve passages that are most relevant to a given query. GraphReader  \cite{li2024graphreader} compresses text into key elements and facts, embedding them into a graph where an agent explores nodes and their connections to gather sufficient information.

In this paper, we focus on GraphRAG \cite{graphrag} and RAPTOR \cite{raptor}, two widely adopted graph-based RAG techniques that have demonstrated robust performance and are representative of current graph-enhanced RAG approaches in practice.

\subsubsection{RAPTOR}
RAPTOR organizes text into a hierarchical tree, applying Gaussian Mixture Models for clustering and iterative summarization. This structure allows efficient retrieval at multiple levels of granularity, ensuring scalability in both build time and token usage \cite{raptor}.

\subsubsection{GraphRAG}
GraphRAG enhances QA by structuring extracted entities and relationships into a knowledge graph. It applies the Leiden algorithm to identify community clusters, generating summaries used as contextual input for answering queries \cite{graphrag}.

Our evaluation focuses on these two techniques as representative graph-based RAG approaches. While other methodologies offer valuable innovations, expanding this comparison remains an avenue for future research.

\section{TREX}\label{sec:overview}

\subsection{Task Setup}
Our task can be formalized by the following. 
Given an open-domain question $q$ and a large body of text $D$, we want to develop a system that can efficiently retrieve the most relevant information from $D$, or a list of text chunks, $\text{Retrieve}(q, D) = \{c_1, c_2, \ldots, c_k\}$, to generate the correct answer $A$ given possible answers $\hat{A}$ to the question $q$. In other words, we want: 
$$
A = \arg \max_{\hat{A}} \mathcal{S}(\hat{A} | q, \text{Retrieve}(q, D) ) 
$$
where $\mathcal{S}$ is a scoring function that evaluates the quality of the answer based on criteria such as relevance, correctness, or user feedback \cite{raptor}. In this work, the scoring function is assessed using a LLM judge, as detailed in Section~\ref{sec:score}.

\subsection{Our Methodology}

Recognizing the limitations of standard vector-based RAG as a standalone solution, we develop a more advanced, cost-efficient method capable of integrating and processing long documents for more reliable responses \cite{liu2024lost}. We recognize the costs associated with certain advanced RAG solutions and prioritize affordability and scalability, ensuring that \sysname remains efficient for large corpora. \sysname is also designed to be query-agnostic, handling both OLTP-style fact-based queries and OLAP-style complex, open-ended questions without requiring an external query router \cite{cosmosaigraph}. Given the range of RAG technologies available for large-scale machine reading and comprehension, we extend the querying modality of RAPTOR, which has demonstrated strong performance across various query types \cite{raptor}. Due to its simple and modular design, \sysname can be incorporated into any RAG system that requires integrating information across documents to improve the accuracy of retrieval and response \cite{microsoft_hybrid_search_ranking}.



\subsubsection{Hierarchical Clustering and Summarization}

\sysname leverages RAPTOR \cite{raptor} to first construct a hierarchical tree structure, $\mathcal{T}$, from the text chunks $D$ using a hierarchical clustering algorithm, represented as $\mathcal{T}(D)$. Each node in this tree is a summary of the child nodes below it. Given a corpus divided into $n$ chunks, the tree will have $n$ leaf nodes, each representing a text chunk $(c_1, c_2, \ldots, c_n)$ \cite{raptor}.

At the first level, we apply a Gaussian Mixture Model (GMM) to cluster text chunks into $k$ clusters. GMMs assume that data arise from a mixture of several Gaussian distributions \cite{mclachlan1988mixture}. Each text chunk $x_i$ is initially represented as a $d$-dimensional vector, derived from embeddings generated by the text-ada-embeddings-002 model~\cite{embedding}. 
The likelihood that a chunk $x_i$ belongs to cluster $k$ is given by $P(x_i | \theta_k) = \frac{1}{(2\pi)^{d/2} |\Sigma_k|^{1/2}} \exp\left(-\frac{1}{2} (x_i - \mu_k)^T \Sigma_k^{-1} (x_i - \mu_k)\right)$, where $\Sigma_k$ and $\mu_k$ represent the covariance matrix and mean vector of cluster $k$, respectively. The overall probability distribution for $P(x)$ is then given by $P(x) = \sum_{k=1}^{K} \pi_k P(x | \theta_k)$, where $\pi_k$ is the weight for that $k^{\text{th}}$ cluster. The GMM model is then trained to maximize the likelihood of the data, $P(x)$ \cite{mclachlan1988mixture, raptor}.

However, text embeddings from the text-ada-embeddings-002 model are high-dimensional ($d$), which can introduce computational inefficiencies and noise. To address this, we apply Uniform Manifold Approximation and Projection (UMAP) as done in \cite{raptor} to reduce the dimensionality from $d$ to $d'$ \cite{mcinnes2018umap}. 
These lower-dimensional representations ($d'$-dimensional vectors) are then used as input to the GMM clustering process. UMAP provides flexible control over neighborhood size, allowing us to capture both global and local structures in the data~\cite{mclachlan1988mixture}.

The optimal number of clusters $k$ is selected automatically using the Bayesian Information Criterion (BIC), which balances model complexity and goodness of fit:
$BIC = \log(n)p-2\log(\hat{L})$ where $\hat{L}$ is the maximized likelihood of the model, $n$ is the number of text chunks, and $p$ represents the number of model parameters, which is a function of $k$. By leveraging BIC, we identify the model with the optimal $k$, ensuring robust clustering performance \cite{raptor}.

\begin{figure}
    \centering
    \includegraphics[width=0.45\textwidth]{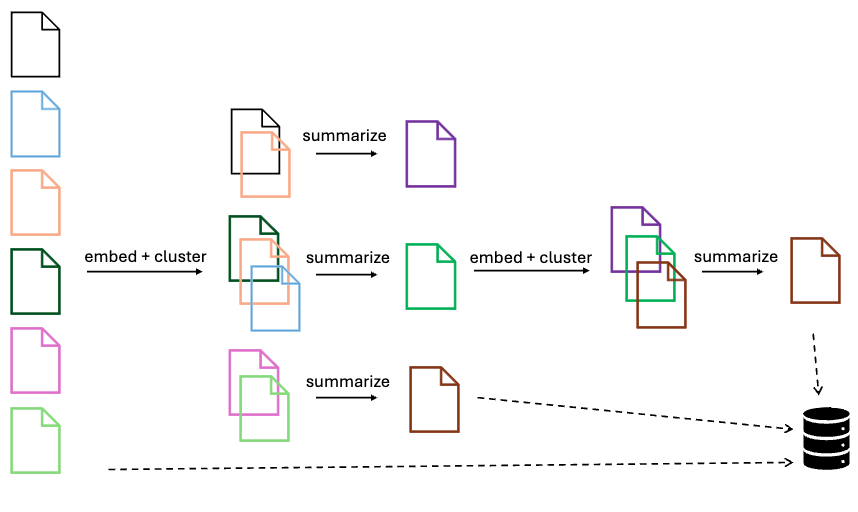}
    \vspace{-.5cm}
    \caption{An example of a hierarchical tree structure built from a set of text chunks ending with the root node. The summary nodes generated by a LLM are then inserted into a vector database \cite{youtube_raptor_long_context}.}
    \label{fig:tree}
\end{figure}

After defining clusters, we use an LLM to generate a summary node for each cluster. At the first level, a summary node is created by $s_i^{l=1} = LLM(c_{i_1}, c_{i_2}, \ldots, 
c_{i_m})$ \eat{\am{are the chunks just concatenated? maybe I would say "a node containing the summary $s_i^l$ is created" or something like that? each s is not only a node but also a chunk of text, right?}} where $m$ is the number of chunks in cluster $i$ at level $l=0$. These summary nodes are subsequently embedded and clustered again, creating higher-level summary nodes that recursively summarize the summaries from the previous level: $s_{j}^{l=2} = LLM(s_{j_1}^{l=1}, s_{j_2}^{l=1}, \ldots, s_{j_m}^{l=1})$ where $m$ is the number of nodes in cluster $j$ at level $l=1$.
This hierarchical process continues until reaching the root node $s_L$, which effectively summarizes the entire document $D$. The resulting hierarchical tree structure is illustrated in Figure~\ref{fig:tree} \cite{raptor}.

\subsubsection{Retrieval and Ranking}
All summary and leaf nodes are stored in a vector database, in our case, we use Azure AI Search Services \cite{microsoft_hybrid_search_ranking}, enabling retrieval of the most relevant nodes in response to user queries. The user query is then encoded using the text-ada-embedding-002 model, and cosine similarity is computed between the query and each summary and leaf node. The top-k nodes are then retrieved and combined with additional nodes from a secondary retrieval mechanism, which uses keyword search on the full text of the query to match terms in the nodes stored in our vector index. The top-k results from this secondary retrieval are then aggregated with the results from the primary retrieval and re-ranked using Reciprocal Rank Fusion (RRF)~\cite{microsoft_hybrid_search_ranking}.

RRF is an unsupervised method that combines rankings from multiple systems, shown to outperform other ranking fusion methods such as Condorcet Fuse and CombMNZ \cite{microsoft_hybrid_search_ranking}. The RRF score for each text chunk, or summary and leaf node, $d$ is calculated as:

$$
\text{RRF} (d \in D) = \sum_{r \in R} \frac{1}{k+r(d)}
$$
where $k$ is a constant (set to the system's default configuration of 60) to reduce the influence of document rank, and $r(d)$ is the rank of document $d$ in the initial set of rankings $R$ from each retreival technique. This method rewards higher-ranked documents while still considering those with lower ranks \cite{microsoft_hybrid_search_ranking}.

For all benchmarking results in Section~\ref{sec:benchmarking}, we use the top five results retrieved by \sysname from the fused ranks of keyword and cosine similarity search as context to generate responses to user queries.

\section{Benchmarking Setup}\label{sec:benchmarking}

We benchmark \sysname alongside GraphRAG and Azure AI Hybrid Search on four publicly available datasets. This selection reflects the range of question types commonly asked by customers, encompassing both the OLTP- and OLAP-style type of queries across diverse topics. The chosen QA datasets include questions that require straightforward responses as well as those demanding synthesis from multiple documents. A summary of our benchmarks is provided in Figure~\ref{fig:spider_map} and described in more detail below.

\subsection{Datasets}
These datasets are all common testbeds for knowledge-intensive tasks and known for their construction of complex questions that require synthesizing information from multiple documents. 

\begin{itemize}
    \item \hotpot: A multi-hop QA dataset from Wikipedia, primarily OLTP-style, requiring synthesis across multiple articles \cite{hotpot}. To mitigate data leakage, we use the \texttt{dev} split (7,405 questions), filtering out queries directly answerable without any context by GPT-4, resulting in 5,491 benchmark questions.
    \item	\msmarco: A large-scale dataset of crowdsourced queries from Bing search logs \cite{msmarco}. It includes over a million queries with annotated, synthesized answers. We sample 1,000 questions for benchmarking.
    \item Microsoft Earnings Call Transcripts: A long-form QA benchmark comprising 40 open-ended financial questions, split between Q4FY2024 transcripts (20 questions derived from a single earnings call) and cross-year corpus queries (20 questions derived from multiple transcripts). All questions demand comprehension and synthesis of earnings calls, whether it be a single or multiple transcripts, to produce multi-sentence answers, emphasizing comprehension over extractive fact retrieval \cite{microsoft_earnings_fy24_q4}.
    \item Kevin Scott Podcast Transcripts: A dataset from GraphRAG, containing 125 high-level questions derived from interviews with tech leaders \cite{scott_behind_the_tech, graphrag}. Unlike standard multi-hop QA, this dataset focuses on multi-document summarization, requiring synthesis of overarching insights rather than retrieval of granular facts.
\end{itemize}

\subsection{Comparative RAG Strategies}
We benchmark \sysname against GraphRAG, RAPTOR, and Azure AI Hybrid Search to assess retrieval and answer generation across OLTP and OLAP-style queries. GraphRAG constructs LLM-generated knowledge graphs to generate community summaries, while RAPTOR organizes text into a hierarchical tree using recursive summarization. Since \sysname integrates graph-based and vector-based RAG techniques, we also compare it against Azure AI Hybrid Search, a vector-based baseline leveraging both keyword and semantic similarity retrieval \cite{microsoft_hybrid_search_ranking}.

\subsubsection{GraphRAG} 
GraphRAG \cite{graphrag} retrieves context via two querying engines:
\begin{itemize}
\item LocalSearch: Optimized for OLTP-style queries by allocating context across entity descriptions, relationships, community summaries, and raw text. Users can fine-tune allocation parameters \cite{graphrag}.
\item GlobalSearch: Optimized for OLAP-style queries using a map-reduce approach that selects and integrates relevant community summaries to provide a holistic response.
\end{itemize}
We use GraphRAG v0.3.6, configuring GlobalSearch with a community hierarchy of 4 and LocalSearch with a hierarchy of 5, allocating 50\% text, 10\% summaries, and 40\% entities/relationships. New indexing mechanisms, such as the use of small language models and traditional NLP techniques, and new querying mechanisms, such as DriftSearch, have recently been introduced \cite{microsoft_graphrag}. A deeper exploration and comparative analysis of these methods is left for future work.

\subsubsection{RAPTOR} RAPTOR structures text as an acyclic hierarchical tree with recursive embeddings and summarization \cite{raptor}. It offers two retrieval strategies:
\begin{itemize}
    \item Tree Traversal: Retrieves top-$k$ similar chunks via cosine similarity and iteratively selects child nodes until reaching a predefined depth or user-defined token limit.
    \item Collapsed Tree Traversal: All nodes are considered simultaneously and the top-$k$ nodes selected based on cosine similarity without hierarchical traversal.
\end{itemize}
We benchmark Tree Traversal with $k=10$ and a max context limit of 3500 tokens \cite{raptor}. 

\subsubsection{Azure AI Hybrid Search} Azure AI Hybrid Search combines keyword and semantic similarity retrieval, fusing results between the two modalalities via Reciprocal Rank Fusion (RRF) to optimize ranking. Additional features include vector field filtering (e.g., time span, category) and custom re-ranking models. We use default settings to evaluate its effectiveness on OLAP and OLTP benchmarks \cite{microsoft_hybrid_search_ranking}.


\subsubsection{Oracle} For \hotpot and \msmarco, we establish an Oracle baseline using annotator-highlighted context, evidence they directly used to answer the query at hand. The gap from 100\% accuracy reflects the LLM's limitations in reasoning and fact synthesis, highlighting areas where the model struggles to fully connect and integrate relevant information \cite{hotpot, msmarco}.

\subsection{Evaluation}
\label{sec:score}
Ensuring the faithfulness and correctness of LLM-generated answers is a key challenge in open-domain QA. A high-quality response should be topical, accurate, and coherent while aligning with the input query and retrieved context. Evaluating OLTP-style queries is straightforward, as they have well-defined ground-truth answers. However, OLAP-style queries, such as ``What is happening in the technology industry?'', introduce subjectivity, as multiple valid responses exist. This variability complicates the definition of correctness and necessitates a more nuanced evaluation approach \cite{graphrag}.

Given the scale of our benchmarks, variability in human judgments, and time constraints of our domain experts, we rely on LLM-as-a-judge for evaluation. Prior studies have shown 80\%+ agreement between human and LLM assessments in MT-Bench and Chatbot Arena \cite{zheng2023judgingllmasajudgemtbenchchatbot}. Future work will refine this framework to better align with expert grading criteria, improving evaluation consistency.

\subsubsection{LLM-Based Evaluation}

To assess answer quality, we apply LLM-as-a-judge for both OLTP and OLAP-style benchmarks:
\begin{itemize}
    \item OLTP Evaluation: Since answers have clear ground truths, we compare model-generated responses against reference answers, achieving 99\%+ agreement. To ensure consistency, we apply a logit bias, restricting outputs to ``YES'' (correct) or ``NO'' (incorrect).
    \item OLAP Evaluation: As no fixed ground truth exists, we adopt the GraphRAG evaluation framework \cite{graphrag}, assessing responses on:
    \begin{itemize}
        \item Comprehensiveness: Depth and thoroughness of information.
	\item Diversity: Inclusion of multiple perspectives.
	\item Empowerment: How well the answer informs decision-making.
    \end{itemize}
\end{itemize}
In future work, we aim to enhance LLM-based evaluation by enabling claim-level verification, ensuring a more granular and human-aligned assessment while maintaining scalability.

\section{Results}

We study the cost-performance trade-offs between \sysname and the various comparative strategies across the four benchmarks. Our evaluation focuses on the quality of the generated answers and the context retrieved, the cost of indexing, and the overall performance of each strategy. Across all experiments, GPT-4o model version 2024-05-13 is used as the LLM for indexing and querying \cite{openai2024gpt4technicalreport}.

\begin{figure}
    \centering
    \includegraphics[width=0.5\textwidth]{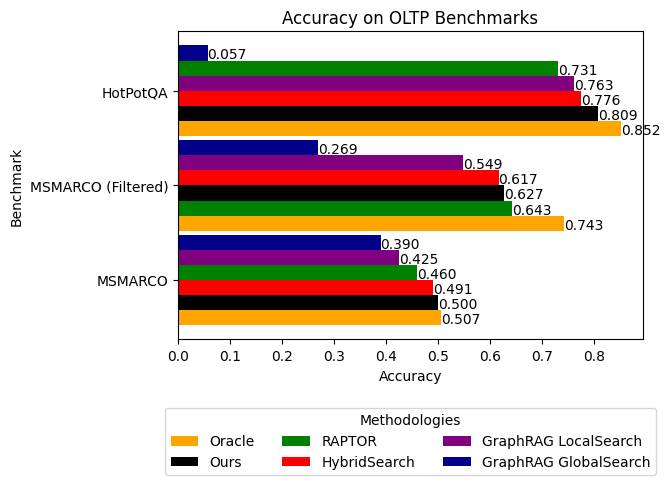}
    \caption{Comparison of accuracy of answers generated by \sysname, GraphRAG, RAPTOR, Azure AI Hybrid Search as well as Oracle on OLTP benchmarks.}
    \label{fig:accuracy}
\end{figure}

\begin{table}[h]
    \centering
    \resizebox{\columnwidth}{!}{
    \begin{tabular}{|l|c|c|c|}
        \hline
        \textbf{Benchmark} & \textbf{GraphRAG} & \textbf{RAPTOR/TREX} & \textbf{HybridSearch} \\
        \hline
        MSMARCO & \$51.37 & \$5.31 & \$0.08 \\
        \hline
        HotPotQA & \$389.12 & \$36.51 & \$0.75 \\
        \hline
        Kevin Scott  & \$63.27 & \$7.03 & \$0.10 \\
        \hline
        Earnings & \$116.71 & \$4.19 & \$0.05 \\
        \hline
    \end{tabular}}
    \caption{Comparison of indexing cost across the four benchmarks on GraphRAG, RAPTOR, \sysname, and HybridSearch methods. Costs are computed based on the use of GPT-4o (\$2.5 per 1 million input and \$10 per 1 million output tokens) and text-ada-embedding-002 (\$0.10 per 1 million tokens) calls. Note that the costs for GraphRAG does not include the use of cost-effective alternatives like small language models for indexing.}
    \label{tab:cost_comparison}
\end{table}

\begin{table}[ht]
    \centering
    \begin{tabular}{|l|c|c|c|c|}
        \hline
        \textbf{Benchmark} & \textbf{GraphRAG} & \textbf{RAPTOR} & \textbf{HybridRAG} & \textbf{TREX} \\
        \hline
        MSMARCO & \$0.04 & \$0.01 & \$0.01 & \$0.01 \\
        \hline
        HotPotQA & \$0.03 & \$0.01 & \$0.01 & \$0.01 \\
        \hline
        Kevin Scott & \$1.41 & \$0.01 & \$0.01 & \$0.01 \\
        \hline
        Earnings & \$0.20 & \$0.10 & \$0.13 & \$0.09 \\
        \hline
    \end{tabular}
    \caption{Comparison of per question average querying costs across the four benchmarks on GraphRAG, RAPTOR, \sysname, and HybridSearch methods. Costs are computed based on the use of GPT-4o (\$2.5 per 1 million input and \$10 per 1 million output tokens) and text-ada-embedding-002 (\$0.10 per 1 million tokens) calls.}
    \label{tab:query_cost_comparison}
\end{table}

\subsection{OLTP-style Benchmarks}

\subsubsection{MSMARCO} In terms of accuracy, \sysname outperforms Graph-RAG, RAPTOR, and Hybrid Search, achieving approximately 50\% accuracy on a sample of 1,000 questions from the MSMARCO benchmark. When filtering out questions without sufficient context---where the top-10 hyperlinks from the Bing search engine do not provide enough relevant information---we are left with a subset of 735 questions. 
The ground truth answers for these filtered questions were labeled by human annotators with variations of the response indicating insufficient context to answer the question.
Typically, \sysname is able to match these non-answers; nevertheless, on the filtered subset of 735, RAPTOR achieves the highest accuracy with 64.3\%, followed closely by \sysname at 62.7\%, Hybrid Search at 61.7\%, and GraphRAG at 54.9\%.

The accuracy results are summarized in Figure~\ref{fig:accuracy}.
In terms of indexing cost, both RAPTOR and \sysname demonstrate savings over GraphRAG as shown in Table~\ref{tab:cost_comparison}. Querying costs are also lower for RAPTOR and \sysname, as indicated in Table~\ref{tab:query_cost_comparison}, suggesting that these methods are better suited for OLTP-style benchmarks.

Given that the MSMARCO benchmark includes annotations of the true evidence used to answer each anonymized query, we report precision and recall metrics based on the retrieved context in the final prompt across the strategies we benchmarked. While substring matching against the ground truth strings provides a basic measure of precision and recall, it is often imprecise for assessing factual correctness due to variations in phrasing and granularity. To address this, we developed a token-based precision and recall method, where the ground truth context is tokenized by converting all text to lowercase and splitting by non-alphanumeric characters. This approach ensures a more granular and flexible evaluation, accounting for variations in wording while maintaining alignment with the essential information needed to answer the query accurately.

As shown in Table~\ref{tab:msmarcoresults}, RAPTOR achieves the highest recall, followed by GraphRAG LocalSearch, HybridSearch, \sysname, and finally GlobalSearch. In terms of precision, RAPTOR also ranks highest, followed by \sysname, HybridSearch, and then GraphRAG. However, precision values are notably low across all methods due to the nature of retrieved context injection---much of the retrieved content is not explicitly included in the ground truth annotations, leading to an apparent drop in precision. This highlights a fundamental limitation in evaluation: ground truth labels often underrepresent the full range of relevant information, making precision an inherently conservative metric. These findings emphasize the trade-off between high recall and precision and the need for retrieval strategies that maximize informative content inclusion while minimizing irrelevant context, ensuring optimal QA performance.

\begin{table}[h]
    \centering
    \begin{minipage}{\columnwidth}
        \centering
        \caption{MSMarco Precision-Recall Results with Standard Deviation in Parentheses}    \label{tab:msmarcoresults}

        \begin{tabular}{|l|c|c|c|}
            \hline
            & \textbf{GlobalSearch} & \textbf{LocalSearch} & \textbf{RAPTOR} \\
            \hline
            Precision & 0.000 (0.00) & 0.002 (0.00) & 0.025 (0.02) \\
            \hline
            Recall & 0.000 (0.00) & 0.577 (0.49) & 0.598 (0.49) \\
            \hline
            Token Precision & 0.000 (0.00) & 0.000 (0.00) & 0.001 (0.00) \\
            \hline
            Token Recall & 0.033 (0.04) & 0.033 (0.04) & 0.033 (0.04) \\
            \hline
        \end{tabular}
    \end{minipage}
    
    \vspace{0.3cm}  

    \begin{minipage}{\columnwidth}
        \centering
        \begin{tabular}{|l|c|c|c|}
            \hline
            & \textbf{HybridSearch} & \textbf{\sysname} \\
            \hline
            Precision & 0.107 (0.11) & 0.103 (0.11) \\
            \hline
            Recall & 0.507 (0.50) & 0.490 (0.50) \\
            \hline
            Token Precision & 0.122 (0.12) & 0.117 (0.11) \\
            \hline
            Token Recall & 0.564 (0.47) & 0.562 (0.47) \\
            \hline
        \end{tabular}
    \end{minipage}

\end{table}

\begin{table}[h]
    \centering
    \begin{minipage}{\columnwidth}
        \centering
        \caption{HotPotQA Precision-Recall Results With Standard Deviation in Parentheses}
        \begin{tabular}{|l|c|c|c|}
            \hline
            & \textbf{GlobalSearch} & \textbf{LocalSearch} & \textbf{RAPTOR} \\
            \hline
            Precision & 0.000 (0.00) & 0.001 (0.00) & 0.002 (0.00) \\
            \hline
            Recall & 0.619 (0.27) & 0.988 (0.082) & 0.904 (0.21) \\
            \hline
        \end{tabular}
    \end{minipage}
    
    \vspace{0.3cm}  

    \begin{minipage}{\columnwidth}
        \centering
        \begin{tabular}{|l|c|c|}
            \hline
            & \textbf{HybridSearch} & \textbf{\sysname} \\
            \hline
            Precision & 0.004 (0.00) & 0.004 (0.00) \\
            \hline
            Recall & 0.988 (0.07) & 0.988 (0.07) \\
            \hline
        \end{tabular}
    \end{minipage}
    \label{tab:hotpotqa_results}
\end{table}

\subsubsection{\hotpot} On the dev split of the \hotpot benchmark, \sysname achieves an accuracy of 80.9\% on a filtered set of 5,491 questions, outperforming GraphRAG, RAPTOR, and Hybrid Search. This accuracy approaches the Oracle benchmark of 85.2\%, which assumes ideal context. In terms of cost, RAPTOR and \sysname offer savings over GraphRAG, with approximately a 10x reduction in indexing costs when using GPT-4o, as shown in Table~\ref{tab:cost_comparison}. Querying costs are also generally lower for RAPTOR and \sysname, as summarized in Table~\ref{tab:query_cost_comparison}.

Precision and recall are calculated here based on the ground truth entities specified in Wikipedia that are necessary to answer each query. Given that ground truth entities are expected to appear exactly as they do in the ground truth context, token precision and recall are not calculated. Using substring matching, recall is highest for GraphRAG LocalSearch, followed by \sysname, Hybrid Search, RAPTOR, and finally GraphRAG GlobalSearch. Precision is highest for \sysname and Hybrid Search, followed by RAPTOR and then GraphRAG. These results suggest that RAPTOR and \sysname may be better suited for OLTP-style benchmarks.

\begin{figure*}[ht]
    \centering
\includegraphics[width=1\textwidth]{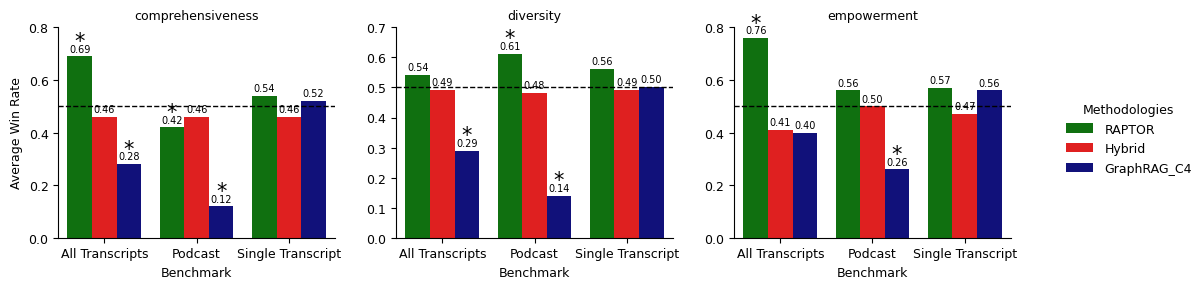} 
    \caption{Mean Win Rates of \sysname versus RAPTOR, GraphRAG Global Search, and Hybrid Search across the metric of comprehensiveness, diversity and empowerment on the two OLAP-style benchmarks. Results that are significant are marked with an asterisk. Global Search results shown are obtained from setting community level at 4 as that hierarchy resulted in the optimal response.}
    \label{fig:mean_win_rates}
\end{figure*}

\subsection{OLAP-style Benchmarks}
\label{sec:olap}

\subsubsection{Microsoft Earnings Call Transcripts} This 40-question benchmark evaluates the performance of the existing suite of RAG strategies by separately analyzing queries sourced from a single earnings transcript versus those requiring synthesis across multiple transcripts. To assess performance, we employ an automated evaluation suite from \citet{graphrag}, using GPT-4o as a judge. This evaluation compares \sysname, RAPTOR, GraphRAG Global Search, and Hybrid Search on metrics of comprehensiveness, diversity, and empowerment. Mean win rate serves as the evaluation metric, with the LLM-as-judge run six times per question and averaged across all comparisons.

As shown in Figure~\ref{fig:mean_win_rates}, \sysname outperforms RAPTOR across all three metrics for both single-transcript and multi-transcript queries. 
Its performance is comparable to Hybrid Search, while its results against GraphRAG Global Search are mixed. For multi-document (multi-transcript) queries spanning the entire corpus of earnings transcripts, GraphRAG excels, producing answers with higher comprehensiveness, diversity, and empowerment. For single-document queries, \sysname seems to slightly outperform GraphRAG in terms of comprehensiveness and empowerment, though the difference is not statistically significant.

Closer analysis reveals that some of the degradation in performance for GraphRAG in the single-document scenario is due to the hallucinations that occasionally arise, as illustrated in the Appendix~\ref{sec:hallucination_sample}. These findings suggest that GraphRAG is particularly effective for OLAP-style questions requiring synthesis across multiple documents, while \sysname provides a cost-effective alternative with moderate performance trade-offs and stronger results for single-document queries. The hierarchical structure of earnings calls, which reflects the natural organization of enterprise topics, aligns well with \sysname's tree-based approach for single transcripts \cite{tan2024htmlrag}. However, this structure may not generalize effectively when handling queries across multiple years of earnings transcripts. Given this, the higher quality responses provided by GraphRAG may justify its increased execution cost as shown in Figure~\ref{tab:cost_comparison}.

\vspace{-.2cm}
\subsubsection{Kevin Scott Podcasts} The Kevin Scott Podcasts benchmark presents a unique challenge due to its open-ended questions, which require detailed and nuanced responses rather than concise answers. Examples of such questions include ``How frequently do discussions about artificial intelligence
arise compared to other topics?'' and ``Which technological sectors do guests believe have the most untapped potential?'' On this benchmark, \sysname outperforms RAPTOR in diversity and empowerment, achieving higher scores 61\% and 56\% of the time, respectively, but falls behind on comprehensiveness. Although \sysname appears to marginally outperform Hybrid Search, the difference is not statistically significant. GraphRAG, however, demonstrates the strongest performance across all metrics.

The thematic and global nature of questions in the Kevin Scott Podcasts dataset aligns well with GraphRAG's graph-structured representation, enabling it to generate highly comprehensive responses. As shown in Appendix~\ref{fig:kevin_scott_comparison}, the results from \sysname are narrower in scope and less prone to hallucinations, while GraphRAG provides more expansive and detailed answers. The performance from GraphRAG on this benchmark is achieved by generating community summaries at a Leiden hierarchy level of 4 and indexing the Kevin Scott corpus at a granularity of 600 tokens. The 600-token granularity reflects how the full corpus is chunked as input into GraphRAG, influencing both the retrieval granularity and the summarization depth. This chunking strategy allows GraphRAG to capture broader thematic connections across the dataset while maintaining a structured representation for multi-document synthesis.









Across the assessed benchmarks, \sysname and RAPTOR demonstrate strong performance for OLTP-style querying. For OLAP-style questions---where answers require significant pre-processing and integration of information from multiple texts---the results are more mixed, as highlighted by our LLM-based evaluation. We evaluate answers on OLAP-style benchmarks using metrics such as comprehensiveness, diversity, and empowerment. While \sysname, RAPTOR, and Hybrid Search produce answers with fewer claims, GraphRAG consistently outperforms these approaches, delivering more comprehensive and expansive responses. However, for OLAP-style queries limited to a single document, \sysname, RAPTOR, and Hybrid Search may be better suited, providing concise and targeted answers.

Future work includes expanding this evaluation platform to assess the correctness and faithfulness of each claim, aiming to replicate the rigor of manual grading for OLAP-style benchmarks.

\begin{figure}
    \centering
    \includegraphics[width=0.45\textwidth]{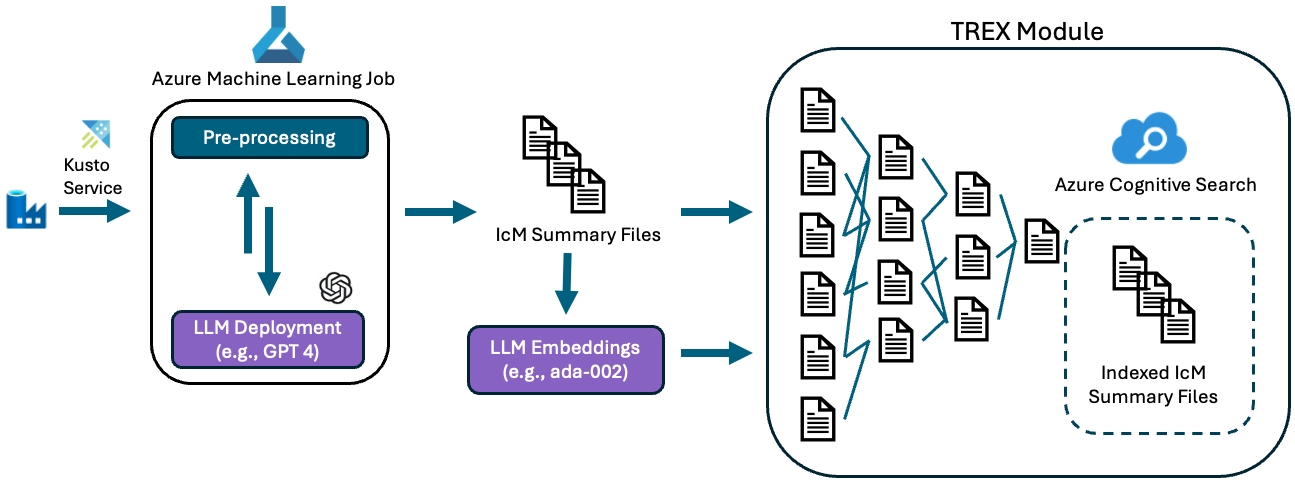}
            
    \caption{Architecture of data preparation system in Azure Data Copilot and how \sysname can be applied to improve contextual retrieval.}
    \label{fig:architeture}
\end{figure}

\section{Case Study}\label{sec:result}

Incident resolution is a manual, time-intensive process across many organizations. Designated responsible individuals (DRIs) work around the clock to address service outages, performance degradations, and other software or hardware failures \cite{chen2020towards}. Developing expertise in incident resolution is challenging, as critical knowledge is often dispersed across siloed repositories, including internal wikis, engineering hubs, and personal SharePoint sites \cite{roy2024exploring}.

To streamline this process, 
an internal system known as 
the Data DRI Copilot \cite{dricopilot,williampaper}, a system that retrieves relevant information from diverse sources to assist engineers in resolving incidents efficiently. Since its launch in September 2023, the system has gained 1,000+ users, processed 30,000+ messages, and maintains an average engagement rating of 3.5/5. While effective for recurring issues and daily tasks, it struggles with novel incidents, where extracting coherent and relevant data from heterogeneous sources is challenging due to inconsistencies in format, terminology, and context. Additionally, specialized code snippets and complex data formats in DRI documents complicate information extraction, occasionally leading to inaccurate responses that hinder decision-making \cite{dricopilot}.

To improve recall and synthesis of information across multiple sources, we evaluate \sysname alongside Azure AI Hybrid Search, assessing its potential to enhance response quality for complex, multi-document queries.


\subsection{Evaluation}
Since launch, DRI Copilot \cite{dricopilot} has received 700+ star ratings, with conversations rated 4 or 5 stars serving as ground-truth data. We evaluate \sysname on a subset of these high-rated question-answer pairs, integrating it into the existing data pipeline (Figure~\ref{fig:architeture}) and comparing its performance against Hybrid Search.

Hybrid Search achieves 39.1\% accuracy, limited by its inability to capture intra- and inter-issue relationships \cite{xu2024retrieval}. In contrast, \sysname improves accuracy to 65.2\%, leveraging multi-hop retrieval structures that mitigate context fragmentation issues common in standard RAG systems.

Future work includes integrating GraphRAG into the pipeline, as its graph-based representation could better capture hyperlinked technical support documents and community clusters, further improving retrieval accuracy and context synthesis in DRI Copilot.

\section{Conclusion}\label{sec:conclude}
This work highlights the need for empirically grounded best practices in Retrieval-Augmented Generation (RAG) and context retrieval mechanisms for LLMs, as guidelines for selecting and optimizing these methods remain largely undocumented. The effectiveness of different RAG strategies---whether vector-based, graph-based, or hybrid approaches---is often unknown until thorough empirical evaluation is conducted. By benchmarking diverse methodologies, this work clarifies when and how various retrieval techniques are most effective, providing insights into their trade-offs across different query types and datasets. As LLMs evolve in an increasingly agentic landscape, their ability to leverage structured knowledge, synthesize multi-source information, and optimize orchestration workflows remains critical. Future research should focus on refining these methodologies and establishing clearer guidelines for integrating LLMs with complex knowledge retrieval systems.
This study identifies where specific graph-based and advanced RAG technologies excel across OLTP- and OLAP-style queries, offering best practices for their deployment. We introduce \sysname, a cost-effective method that handles both query types with minimal trade-offs, while emphasizing the need for rigorous evaluation standards to drive more effective adoption of AI-driven retrieval systems.

\section*{Acknowledgments}
We extend our gratitude to Aleksey Savateyev and Chris Joakim for their insights on CosmosAIGraph. 
We also thank the members of the GraphRAG team---Alonso Fernandez, Andres Esquivel, Chris Trevino, Darren Edge, Dasha Metropolitansky, Nathan Evans and Robert Ness---for their support in auditing our work and sharing best practices for deploying GraphRAG effectively. Special thanks to Harsha Simhadri for pointing us towards the MSMARCO benchmark and Avrilia Floratou, Brice Chung, Sandip Kulkarni, and Aydan Aksoylar for motivating this project.

\bibliographystyle{ACM-Reference-Format}

\appendix  

\section{Additional Results}  
\label{sec:appendix}

Please note that hallucinations from the LLM are ascribed in blue. 

\subsection{Example LLM-generated answers using the GlobalSearch modality in GraphRAG \cite{graphrag}} 
\label{sec:hallucination_sample}
\textbf{Question}: As per Satya, what initiatives has Microsoft taken to enhance security? 

            \textbf{Overview} \\ Under the leadership of Satya Nadella, Microsoft has taken significant steps to enhance security across its various platforms and services. These initiatives are part of a broader strategy to ensure that security is integrated by design, by default, and throughout the operational lifecycle of its products.

            \textbf{Secure Future Initiative} \\ One of the key security initiatives under Satya Nadella is the "Secure Future Initiative." This initiative emphasizes the principles of being secure by design, default, and in operations. It reflects Microsoft's commitment to building security into the fabric of its software and services from the ground up [Data: Entities (138)].

            \textbf{Integration of Security in Microsoft Products} \\
            Microsoft's approach to security is not limited to specific products but spans across its entire ecosystem. This includes integrating advanced security features into its flagship products like Microsoft 365, Azure, and its various SaaS offerings. The integration of security features is designed to protect against a wide range of threats, ensuring that both consumer and enterprise data are safeguarded.

            \textcolor{blue}{\textbf{Focus on AI and Security} \\
            In addition to traditional security measures, Microsoft under Nadella's leadership is focusing on leveraging artificial intelligence (AI) to enhance security protocols. This involves using AI to predict and prevent security breaches before they occur, thereby adding an additional layer of security that is proactive rather than reactive.}

            \textbf{Conclusion} \\ 
            The security initiatives led by Satya Nadella are a testament to Microsoft's commitment to safeguarding its products and services against emerging threats. By embedding security deeply into the architecture of its offerings and using cutting-edge technologies like AI, Microsoft aims to stay ahead in the cybersecurity domain, ensuring trust and reliability for its users worldwide \cite{microsoft_earnings_fy24_q4, graphrag}.

\subsection{Example of an entity generated from the LLM-index graph using default settings from the GraphRAG accelerator \cite{graphrag}}

 \textbf{JSON Blob:}

            \{
                \newline
                \hspace*{0.5cm} "id": "181",
                \newline
                \hspace*{0.5cm} "entity": "AMY",
                \newline
                \hspace*{0.5cm} "description": "Amy is a high-ranking official or executive at Microsoft, likely involved in financial and strategic discussions at the company. She is actively engaged in conversations about capital expenditure, particularly focusing on Azure revenue growth and its impact on capex. \textcolor{blue}{Amy's expertise also extends to Dynamics, indicating her involvement in business operations.} She participates in financial calls, often addressing questions alongside Satya Nadella, and contributes valuable insights into asset management strategies. Additionally, Amy discusses cloud margins and AI investments, analyzing their margin impact both in the short term and long term. Her role allows her to add significant information to discussions about capital expenditure investments and yields, highlighting her comprehensive understanding of Microsoft's financial strategies.",
                \newline
                \hspace*{0.5cm} "number of relationships": "10"
                \newline
            \}

\subsection{LLM-generated answer to OLAP query} 
\label{fig:kevin_scott_comparison} 

Comparison of answers from GraphRAG GlobalSearch and \sysname in response to the query ``How do tech leaders describe the influence of technology on everyday life?'' in the Kevin Scott podcast benchmark \cite{graphrag, raptor}.

\subsubsection{Answer from GraphRAG GlobalSearch}
\noindent\hrule
\vspace{0.1cm}
\textbf{The Profound Impact of Technology on Everyday Life}

Tech leaders consistently highlight the transformative influence of technology on various aspects of everyday life. The Internet, for instance, has revolutionized how we access information, communicate, and collaborate, serving as a pivotal resource for innovation across multiple fields such as marketing, social media, AI, and music production [Data: Reports (871)]. Bill Gates emphasizes the role of personal computing and the Internet in making technology more accessible and advancing its capabilities, which has had a profound impact on modern society [Data: Reports (870)].

\textbf{Artificial Intelligence and Automation}

Artificial Intelligence (AI) is frequently cited as a transformative technology with significant potential for future advancements. Kevin Scott and other tech leaders underscore AI's role in enhancing productivity, streamlining operations, and addressing significant challenges in various domains, including healthcare, agriculture, and education [Data: Reports (58, 553, 555, 357, 543, +more)]. Automation, powered by AI, is transforming industries by reducing the need for human intervention in repetitive tasks, thereby enhancing efficiency and productivity [Data: Reports (809)].

\textbf{Ethical Considerations and Societal Impact}

The ethical implications of technological advancements are a recurring theme among tech leaders. Figures like Reid Hoffman and Judy Estrin emphasize the importance of responsible AI development and the need for ethical frameworks to guide the deployment of new technologies. These discussions highlight concerns about potential biases, privacy issues, and the broader societal impacts of AI and other technologies [Data: Reports (891, 689, 692, 377, +more)].

\textbf{Technological Advancements in Healthcare}

AI's integration into healthcare is another significant area of impact. Eric Horvitz and other leaders discuss how AI can enhance medical practices, improve patient outcomes, and transform healthcare delivery. Technologies such as deep learning algorithms and smart sensors are being utilized in settings like ICUs and senior homes to improve care quality [Data: Reports (880, 510, 543, 878, +more)].

\textbf{Everyday Technologies and Consumer Impact}

Consumer technologies like smartphones, IoT devices, and augmented reality (AR) have become integral to daily life. Smartphones, for example, combine mobile telephone and computing functions, making computing accessible and affordable to the general public [Data: Reports (58)]. IoT devices like Alexa and Google Home provide convenience and automation, enhancing everyday interactions [Data: Reports (588)]. AR technologies enrich real-world environments with digital information, offering interactive and engaging experiences [Data: Reports (776)].

\textbf{The Role of Technology in Education and Social Justice}

Tech leaders also highlight the role of technology in education and social justice. Initiatives like Coursera, led by Andrew Ng, have democratized education, making learning more accessible worldwide [Data: Reports (498)]. Danah Boyd's work emphasizes the impact of technology on social justice, literacy, and community engagement, reflecting broader concerns about the societal implications of technological advancements [Data: Reports (377, 378)].

\textbf{Conclusion}

In summary, tech leaders describe the influence of technology on everyday life as profound and multifaceted. From enhancing productivity and healthcare to addressing ethical concerns and promoting social justice, technology continues to shape and redefine various aspects of our daily experiences. The ongoing advancements in AI, automation, and consumer technologies underscore the transformative potential of technology in improving quality of life and addressing global challenges.

\subsubsection{Answer from \sysname} 
\noindent\hrule
\vspace{0.1cm}

\textbf{Influence of Technology on Everyday Life}

Tech leaders often highlight the transformative potential of technology and its profound impact on various aspects of everyday life. This influence spans multiple domains, including education, healthcare, and community dynamics, among others.

\textbf{Education and Learning}

Bill Gates emphasizes the significant role technology plays in revolutionizing education. He discusses how accessible technology can enhance corporate learning activities by allowing tasks to be programmed using natural language. This not only streamlines processes but also engages users more effectively. Gates finds education to be one of the most exciting applications of new technology, noting its potential to make learning more efficient and widespread [Data: Reports (2, 4, +more)].

\textbf{Healthcare and Well-being}

The integration of technology in healthcare is another area where its influence is markedly beneficial. For instance, advanced AI platforms are being developed to help individuals improve their lives by making healthier and more productive choices. The potential for AI to act as a natural human-machine interface, gleaning information from everyday interactions and nudging individuals towards better habits, is discussed extensively. This includes the ethical considerations of how technology should influence behavior without crossing into manipulation [Data: Reports (7, +more)].

\textbf{Community and Job Repatriation}

Technology also plays a pivotal role in community development and job creation. The repatriation of jobs to local communities, previously outsourced overseas, is facilitated by the use of advanced automated machines. These machines, driven by the principles akin to Moore's Law, provide more value per dollar spent, making local businesses competitive. This technological leverage not only boosts local economies but also inspires community members by bringing back jobs and fostering a sense of pride and motivation [Data: Reports (1, +more)].

\textbf{Ethical Considerations and Influence}

The ethical implications of technology's influence are a recurring theme among tech leaders. There is a consensus that while technology should assist and enhance human capabilities, it must be designed and deployed ethically. The line between positive behavior nudging and manipulation is thin, and it is crucial to ensure that technology serves the user's best interests without imposing unintended biases or goals [Data: Reports (7, +more)].

\textbf{Conclusion}

In summary, tech leaders describe the influence of technology on everyday life as multifaceted and transformative. From revolutionizing education and healthcare to rejuvenating local economies and addressing ethical challenges, technology holds the promise of significant positive impact. However, it is essential to navigate its deployment thoughtfully to maximize benefits while minimizing potential drawbacks \cite{graphrag, raptor}.

\end{document}